\documentclass[twocolumn]{aastex6}
\usepackage{amsmath}
\usepackage{CJK}

\makeatletter
\newcommand{\bdv}[1]{\mbox{\boldmath$#1$}}

\def\e{{\rm E}}
\def\l{{\rm L}}
\def\s{{\rm S}}
\def\rel{{\rm rel}}

\def\max{{\rm max}}

\def\microas{\mu{\rm as}}

\def\masyr{{\rm mas~yr^{-1}}}
\def\K2C9{{\emph{K2}C9}}
\def\piee{{\pi_{\rm E,E}}}
\def\pien{{\pi_{\rm E,N}}}
\def\kep{{\rm kep}}
\def\bkg{{\rm bkg}}
\def\detr{{\rm detr}}
\def\phot{{\rm phot}}

\shorttitle{\K2C9 Microlensing Photometry}
\shortauthors{Zhu et al.}

\begin{document}
\begin{CJK*}{UTF8}{gbsn}

\title{Extracting Microlensing Signals from \emph{K2} Campaign 9}

\author{
Wei~Zhu~(祝伟)\altaffilmark{1},
C.~X.~Huang~(黄煦)\altaffilmark{2,3,4},
A.~Udalski\altaffilmark{5},
M.~Soares-Furtado\altaffilmark{6,8},
%%%%%%%%
R.~Poleski\altaffilmark{1,5},
J.~Skowron\altaffilmark{5},
P.~Mr{\'o}z\altaffilmark{5},
M.~K.~Szyma{\'n}ski\altaffilmark{5},
I.~Soszy{\'n}ski\altaffilmark{5},
P.~Pietrukowicz\altaffilmark{5},
S.~Koz{\L}owski\altaffilmark{5},
K.~Ulaczyk\altaffilmark{5,7},
M.~Pawlak\altaffilmark{5}}

\email{zhu.908@osu.edu}

\altaffiltext{1}{Department of Astronomy, Ohio State University, 140 W. 18th Ave., Columbus, OH  43210, USA}
\altaffiltext{2}{Department of Physics and Kavli Institute for Astrophysics and Space Research, Massachusetts Institute of Technology, Cambridge, MA 02139, USA}
\altaffiltext{3}{Dunlap Institute for Astronomy and Astrophysics, University of Toronto, Toronto, ON M5S 3H4, Canada}
\altaffiltext{4}{Centre of Planetary Science, University of Toronto, Scarborough Campus
Physical \& Environmental Sciences, Toronto, M1C 1A4, Canada}
\altaffiltext{5}{Warsaw University Observatory, Al. Ujazdowskie 4, 00-478 Warszawa, Poland}
\altaffiltext{6}{Department of Astrophysical Sciences, Princeton University, Princeton, NJ 08544, USA}
\altaffiltext{7}{Department of Physics, University of Warwick, Gibbet Hill Road, Coventry, CV4~7AL, UK}
\altaffiltext{8}{National Science Foundation Graduate Research Fellow}

%\submitted{To be submitted to PASP}

\begin{abstract}
    The reduction of the \emph{K2}'s Campaign 9 (\K2C9) microlensing data is challenging mostly because of the very crowded field and the unstable pointing of the spacecraft. In this work, we present the first method that can extract microlensing signals from this \K2C9 data product. The raw light curves and the astrometric solutions are first derived, using the techniques from Soares-Furtado et al. and Huang et al. for \emph{K2} dense field photometry. We then minimize and remove the systematic effect by performing simultaneous modeling with the microlensing signal. We also derive precise $(K_p-I)$ vs. $(V-I)$ color-color relations that can predict the microlensing source flux in the \emph{Kepler} bandpass. By implementing the color-color relation in the light curve modeling, we show that the microlensing parameters can be better constrained. In the end, we use two example microlensing events, OGLE-2016-BLG-0980 and OGLE-2016-BLG-0940, to test our method.
\end{abstract}


%%%%%%%%%%%%%%%%%%%%%%%%%%%%%%%%%
%\input{sec1.tex}
\section{Introduction} \label{sec:introduction}

One challenge in Galactic microlensing observations is to derive physical parameters of the lens object/system from the observed signal. In particular, three observables are required in order to disentangle the lens (total) mass $M_\l$, the lens distance $D_\l$ and the lens-source relative proper motion $\mu_\rel$: the event timescale $t_\e$, the angular Einstein radius $\theta_\e$, and the microlensing parallax parameter $\bdv{\pi_\e}$ that quantifies the lens-source relative displacement in the unit of $\theta_\e$ \citep{Gould:2000}. 
Of these three observables, $\bdv{\pi_\e}$ is usually the most crucial one. This is because, the timescale $t_\e$ is almost always measured precisely except for extremely long ($t_\e\gtrsim300~$d) or short ($t_\e\lesssim2~$d) events, and $\theta_\e$ is measurable in events that show planetary or binary anomalies \citep{Zhu:2014,Suzuki:2016}. For the majority of single-lens events, for which $\theta_\e$ cannot be measured directly from the light curves, measurements of $\pi_\e$ can still put very tight ($\sim25\%$) constraints on the inferred lens mass and distance \citep{HanGould:1995,Zhu:2017}. 

Simultaneous observations from at least two well-separated ($\sim1$~AU) observatories has always been considered the best way to measure the microlensing parallax \citep{Refsdal:1966,Gould:1994}. The microlensing observations using the \emph{Spitzer} spacecraft in an Earth trailing orbit have been very successful, and yielded important results \citep[e.g.,][]{Dong:2007,Udalski:2015a,Yee:2015a,SCN:2015,Zhu:2015}. However, due to the small field of view, \emph{Spitzer} is only able to follow up the microlensing events found from the ground, and given the routine of selecting and uploading events, \emph{Spitzer} is not able to measure $\bdv{\pi_\e}$ for any events with microlensing event timescale $t_\e\lesssim 4$ days. This precludes \emph{Spitzer} from observing events that are potentially caused by free-floating planets (FFPs), which usually have $t_\e\lesssim2$~days \citep{Sumi:2011}. Another significant disadvantage of the follow-up mode of \emph{Spitzer} microlensing is that it is difficult to quantify the selection effect, unless it is carefully controlled in designing the experiment \citep{Yee:2015b}.

Similar to \emph{Spitzer}, the \emph{Kepler} space telescope \citep{Borucki:2010} is also in heliocentric orbit and thus suitable for microlensing observations. Indeed, Campaign 9 (C9) of the two-wheeled \emph{Kepler} mission (\emph{K2}, \citealt{Howell:2014}), or \K2C9, is dedicated for Galactic microlensing observations \citep{GouldHorne:2013,Henderson:2016}.

However, the reduction of the \K2C9 microlensing data is challenging. In the \emph{K2} era, the pointing of the spacecraft drifts by more than one pixel (4$''$) for every $\sim$6.5 hrs, resulting in a significant systematic effect. This can nevertheless be overcome by introducing specific techniques. For example, \citet{Vanderburg:2014} pointed out that the systematics correlates with the stellar centroid positions. By using the centroids determined by the point source extractor, \citet{Vanderburg:2014} can achieve a photometric precision for isolated \emph{K2} targets that are only factors of $2-3$ worse than the original \emph{Kepler} mission. \citet{Huang:2015} undertook a similar approach, but determined the centroids of target stars by matching isolated reference stars in the \emph{K2} frames to the UCAC4 catalog \citep{Zacharias:2013}.

Methods have also been proposed for \emph{K2} crowded field photometry. \citet{Libralato:2016} applied the PSF (point spread function) neighbor-subtraction method that was well established for other space telescopes to the Campaign 0 cluster stars. While they are able to achieve $\sim$1\% photometric precision on cluster stars with \emph{Kepler} magnitude $K_p=18$, their approach requires careful reconstruction of the \emph{K2} PSF. \citet{MSF:2017} performed the more traditional image subtraction technique on the similar data set, and achieved a similar photometric precision.

The real challenge in reducing \K2C9 microlensing data comes from the fact that the microlensing field is the most crowded in all \emph{K2} fields. In particular, the \K2C9 field is a factor of $\sim$10 denser in stellar number density than the Campaign 0 clusters that were analyzed in \citet{Libralato:2016} and \citet{MSF:2017}. As Figure 8 of \citet{Henderson:2016} illustrates, there is on average one star brighter than $I=18$ on each \emph{Kepler} pixel, and one star brighter than $I=15.5$ on a 10 pixel$^2$ aperture. For a comparison, the typical microlensing sources have $I\approx18$ at the baseline.

In this work, we apply the photometric methods of \citet{Huang:2015} and \citet{MSF:2017} to the \K2C9 microlensing data set, and develop modeling techniques that can properly extract the microlensing signals. Short summaries of \K2C9 project and the photometric methods are given in Section~\ref{sec:method}. The modeling techniques are presented in Section~\ref{sec:modeling}. We then apply our method to two example microlensing events in Section~\ref{sec:examples}, and discuss our method and its implications in Section~\ref{sec:discussion}.

%%%%%%%%%%%%%%%%%%%%%%%%%%%%%%%%%
\section{\K2C9 Data Reduction} \label{sec:method}
\subsection{\K2C9 Description}

We provide here a brief summary of \K2C9 observations, and recommend interested readers to \citet{Henderson:2016} for a detailed description. 

\K2C9 was conducted during 2016 April 22
\footnote{The nominal start date of \K2C9 was 2016 April 7. However, the spacecraft entered into Emergency Mode prior to this intended start date, and the full recovery led to this actual start data.}
and 2016 June 2. Unlike most of other \emph{K2} campaigns, this campaign was divided into two sub-campaigns, with April 22 to May 18 for C9a and May 22 to June 2 for C9b, and data from C9a were downloaded during the mid-campaign break (May 19--21, or 7527.4$<$JD-2450000$<7531.0$), in order to achieve total super stamp area as large as 3.7 deg$^2$. The five super stamps are observed on two different CCD modules, and read out in channels 30, 31, 32, 49 and 52.   
The sky region covered by the super stamp, centered at approximately (R.A., decl.)$_{2000}=(17^{\rm h}57^{\rm m},\ -28\arcdeg24\arcmin)$, was chosen to maximize the total microlensing event rate \citep{Poleski:2016} as well as to accommodate ground-based survey observations. See Figure~7 of \citet{Henderson:2016} for the positions of \K2C9 and the microlensing super stamp.
In addition, microlensing events that were identified from the ground and expected to be observable by \emph{Kepler} were added as \emph{late targets}, and a postage stamp with size the square of a few hundred pixels was added for each late target. In total, 70 unique late targets were observed in this way. All observations were in the long cadence (30 minutes) mode.

Together with simultaneous observations from the ground,
\footnote{See Figure~9 and Table~3 of \citet{Henderson:2016} for the ground-based observing resources concurrent with \K2C9.}
\K2C9 is expected to yield microlensing parallax measurements for nearly 200 events. Simulations show that \K2C9 can detect up to 10 short timescale events, including one with finite-source effect \citep{Yoo:2004} that will enable a direct mass measurement \citep{Penny:2016}. The survey mode of \K2C9 also makes it easy to model the detection efficiency, and thus possible to study statistically the mass function of other extremely faint or even dark objects as well as to combine with the \emph{Spitzer} microlensing sample in order to refine the Galactic distribution of planets \citep{SCN:2015,Zhu:2017}.

\begin{figure*}
\epsscale{1.2}
\plotone{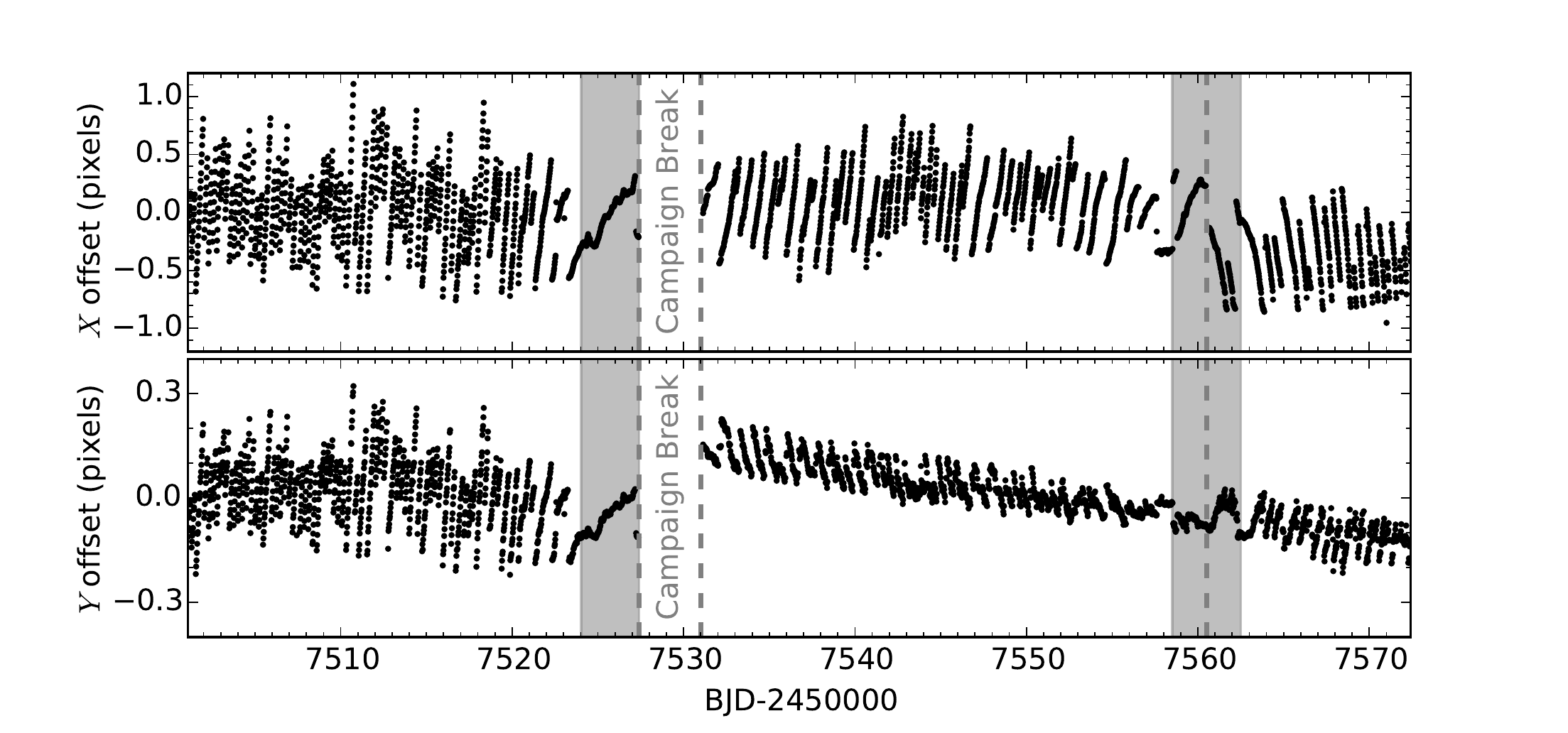}
\caption{The pointing drift pattern (in $X$ and $Y$ directions) of \emph{K2} module 31 during the campaign 9. No data were taken during the mid-campaign break ($7527.4-7531.0$). Together with the change in the drift direction at 7560.5, the light curve from campaign 9 is divided into three segments. The two broad gray bands indicate the window within which the spacecraft underwent irregular drifts (due to the commanded pointing offset for C9b and an impulsive roll offset in C9b, see data release note for \K2C9). Data taken during these two windows can be used in the light curve modeling, but any anomalous feature occurred within them should be taken with caution.
\label{fig:driftings}}
\end{figure*}

\subsection{Differential Photometry on \K2C9 data}

We acquire the \K2C9 target pixel files (TPFs) through NASA's Barbara A. Mikulski Archive for Space Telescopes (MAST)
\footnote{\url{https://archive.stsci.edu/k2/}},
and then stick all TPFs together into ``Sparse Full Frame Images'' (SFFIs). In order to ensure a proper reconstruction, these SFFIs are cross-checked with the full frame images downloaded for Campaign 9.

We then perform the differential photometry on the \K2C9 data following the methods described in \citet{MSF:2017} and \citet{Huang:2015}. This is done for C9a and C9b separately.
We first find the astrometric solution of individual frame, by matching {a large number ($\sim$1000) of} bright ($R<16$) stars of the UCAC4 catalog \citep{Zacharias:2013} to the point sources found in the frame. This astrometric information describes the pointing drift pattern of the spacecraft, and is used in both the photometry extraction and the light curve modeling. For each \emph{K2} module, this pointing drift pattern is defined as the $X$ and $Y$ positions of the center pixel at each epoch. As an example, we show the pointing pattern of channel 31\ in Figure~\ref{fig:driftings}.

We use the \texttt{FITSH} package \citep{Pal:2012} for the image subtractions. This package implements the classic method of \citet{AlardLupton:1998}, and is developed for reducing images from the HATNet project \citep{Bakos:2004}.
We first select a frame that has the median drift as the astrometric reference frame. Then, we select another 20 frames that have approximately the same (offset$<0.15$ pixel) pointing as the astrometric reference frame, and take the median as the master photometric reference frame. 
We adopt the convolution kernel from \citet{MSF:2017} to create the {difference} images between individual frames and the master photometric reference frame. This kernel, which is a discrete kernel with half-size of two pixels and no spatial variation across the frame, is shown to produce the best photometry for crowded fields observed in \emph{K2} campaign 0.

Finally, we extract \emph{raw light curves} from the {difference} images and the master photometric reference image, using the centroids that are determined by the astrometry solution. This is done for 40 circular apertures with radii ranging from 2.5 to 5.5 pixels. For each target, the best aperture is chosen to be the one with the minimum scattering after the detrend procedure. {The raw light curves of two selected microlensing events are shown in the upper panels of Figures~\ref{fig:0980-kepler} and \ref{fig:0940-kepler}.}

%%%%%%%%%%%%%%%%%%%%%%%%%%%%%%%%%
\section{Detrending \& Microlensing Modeling} \label{sec:modeling}

\subsection{Methodology} \label{sec:methodology}

As earlier studies show \citep[e.g.,][]{Vanderburg:2014}, the \emph{K2} raw light curves show strong systematic effects that correlate with the spacecraft motion. The detrending process is therefore performed, yielding significantly improved photometry while preserving the physical signals. In the transit case, the detrending process can be done separately from the light curve modeling, because the transit signal lasts for $\lesssim1~$day. However, this is nevertheless not the case for microlensing signals, which typical last tens of days, and can not always be traced by a low order spline. The External Parameter Decorrelation (EPD) technique that is proposed by \citet{Huang:2015} to preserve long-term trend in the raw light curve does not work well either. In almost all cases ({$\sim$10 events including the two demonstrated in Section~\ref{sec:examples}}) that we have checked, this technique either destroys the microlensing signal or distorts the signal significantly. 

Instead, we perform a simultaneous modeling of the microlensing signal and the systematic effects. For a given \K2C9 light curve segment $j$, the total flux at time $t_i$ is modeled as
\begin{equation} \label{eqn:model-kepler}
\begin{split}
    F^\kep(t_i) & = F_\s^\kep \cdot {(A_i^\kep-1)} + c_0^j + c_1^j F^{\bkg,j}_i + c_2^j\sigma(F^{\bkg,j}_i) \\
                & \quad + \sum_{k=1}^{N_\detr} \Big[ c_{k,1}^j\sin(2\pi k X_i) + c_{k,2}^j\cos(2\pi k X_i) \\
                & \quad + c_{k,3}\sin(2\pi k Y_i) + c_{k,4} \cos(2\pi k Y_i) \Big] \ .
\end{split}
\end{equation}
Here $F_\s^{\rm kep}$ is the {unmagnified} flux of the microlensing source in the \emph{Kepler} band, $A_i^{\rm kep}$ is the source brightness magnification as seen by \emph{Kepler} at given $t_i$, $\{c_0,c_1,c_2,c_{k,1},c_{k,2},c_{k,3},c_{k,4}\}$ are the detrending parameters, $F^{\rm bkg}_i$ and $\sigma(F^{\rm bkg}_i)$ are the background flux {(i.e., blend flux in microlensing term)} and its uncertainty, and $X_i$ and $Y_i$ are the relative drift of the module at $t_i$ along two directions, respectively. Unlike $F_\s^{\rm kep}$, $A_i$ and $\{c_0,c_1,c_2,c_{k,1},c_{k,2},c_{k,3},c_{k,4}\}$, parameters $F^{\rm bkg}_i$, $\sigma(F^{\rm bkg}_i)$, $X_i$, and $Y_i$ are known for any given $t_i$ based on the photometry and astrometry results from Section~\ref{sec:method}.
The $\sin$ and $\cos$ terms (a.k.a., detrending terms) in Equation~(\ref{eqn:model-kepler}) account for the inter-pixel and intra-pixel sensitivity variations. A larger $N_\detr$ is in general more efficient at removing trends that are shorter than 6.5~hrs, which in our case are essentially systematic. However, including high-order terms also has the risk of over-fitting short-term trends while under-fitting the long-term trend. For our data, we find $N_\detr\le5$ is reasonable, and suggest to use the upper bound unless it fails catastrophically. Nevertheless, the resulting microlensing parameters are not very sensitive to the choice of $N_\detr$. See Section~\ref{sec:examples} for example events and relevant discussions.

%\replaced{It is crucial to include ground-based data in the microlensing modeling, in order to yield better constrained as well as self-consistent microlensing parameters. For space-based microlensing observations, the microlensing parallax parameter can be approximated as
%\begin{equation}
%    \bdv{\pi_\e} \approx \left(\frac{\Delta t_0}{t_\e},\Delta u_0\right)\ ,
%\end{equation}
%where $\Delta t_0$ and $\Delta u_0$ are the differences in event peak time $t_0$ and impact parameter $u_0$ as seen from the satellite and the ground.
%However, an accurate and self-consistent solution has to be derived by simultaneously modeling the space-based and ground-based data, which takes into account the actual motion of the satellite
%\footnote{This can be acquired from the JPL Horizons website: \url{http://ssd.jpl.nasa.gov/?horizons}.}.
%Because the detrending of the \K2C9 light curve is model-dependent, it is likely to yield inconsistent result if one models the \K2C9 light curve solely and then measures $\bdv{\pi_\e}$ jointly with ground-based data.}
%%%%%%%%%
{It is crucial to include ground-based data in the microlensing modeling, in order to yield better constrained as well as self-consistent microlensing parameters. This is because, although the microlensing parallax can be estimated via the differences in event peak time $t_0$ and impact parameter $u_0$ as seen from the satellite and the ground \citep{Gould:1994}
\begin{equation} \label{eqn:pie-approx}
    \bdv{\pi_\e} \approx \left(\frac{\Delta t_0}{t_\e},\Delta u_0\right)\ ,
\end{equation}
an accurate and self-consistent solution has to be derived by simultaneously modeling the space-based and ground-based light curves and taking into account the geocentric motion of the satellite
\footnote{This can be acquired from the JPL Horizons website: \url{http://ssd.jpl.nasa.gov/?horizons}. Note that the satellite-Earth relative motion can be ignored only if this relative velocity is considerably small compared to the lens-source projected velocity. See \citet{ZhuGould:2016} for a detailed discussion.}.
Therefore, we choose to model all the variable terms simultaneously, including the microlensing signal in the ground-based data, the microlensing signal in \K2C9 data, and the systematics in \K2C9 data. Another reason for doing so is that, the detrending of the \K2C9 light curve can be model-dependent, especially for events with partial coverage and/or long timescales. In such cases, inconsistent results may appear if one models the \K2C9 light curve solely and them measures $\bdv{\pi_\e}$ jointly with ground-based data.
}

For the ground-based data sets, the total flux at time $t_i$ for data set $j$ is modeled by
\begin{equation}
    F^j (t_i) = F_\s^j \cdot A_i^j + F_{\rm B}^j\ ,
\end{equation}
where $F_\s$ is the source flux at baseline. The parameter $F_{\rm B}$ accounts for the flux that is within the aperture but does not participate in the event. Unlike the case of \K2C9 (Equation~\ref{eqn:model-kepler}), the blending flux remains constant throughout all observations. Different ground-based data sets are assigned with different flux parameters $F_\s$ and $F_{\rm B}$.

There have been many methods and algorithms proposed to compute the microlensing magnification $A$ for given microlensing parameters, for single-lens events \citep[e.g.,][]{WittMao:1994,Yoo:2004} and multiple-lens events \citep[e.g.,][]{Dong:2006,Bozza:2010}. For the purpose of this paper, we only consider the relatively simple single-lens events. The magnification $A$ can be computed from a set of microlensing parameters $\{t_0,u_0,t_\e,\rho,\piee,\pien\}$: $t_0$ is the peak of the event as seen from the ground, $u_0$ is the impact parameter normalized to the angular Einstein radius $\theta_\e$, $t_\e$ is the event timescale, $\rho$ is the source angular radius normalized to $\theta_\e$, and $\piee$ and $\pien$ are components of the microlensing parallax vector $\bdv{\pi_\e}$ along the east and north direction, respectively. Note that parameters $t_0$, $u_0$ and $t_\e$ are defined in the geocentric frame \citep{Gould:2004}.

To summarize the degrees of freedom in our model, we have in total tens of free parameters: a set of microlensing parameters, which are $\{t_0,u_0,t_\e,\rho,\piee,\pien\}$ in the single-lens case,
\footnote{Note that for the majority of single-lens events, there is no the finite-source effect to constrain $\rho$ \citep{Zhu:2016}, and therefore $\rho$ is usually a fixed parameter.}
shared by all data sets; different flux parameters $F_\s$ and $F_{\rm B}$ for different ground-based data sets; one common source flux parameter $F_\s^{\rm kep}$ for all \K2C9 light curve segments; and different detrending parameters $\{c_0,c_1,c_2,c_{k,1},c_{k,2},c_{k,3},c_{k,4}\}$ for {the three} different \K2C9 light curve segments. We employ the Markov Chain Monte Carlo (MCMC) analysis to search for the best solution and estimate the uncertainties of parameters. In this MCMC analysis, we only have the microlensing parameters and $F_\s^{\rm kep}$ as free chain parameters, and derive the best values of all other parameters analytically via maximum likelihood estimation \citep[e.g.,][]{Ivezic:2014}. This has negligible effects on the best-fit solution as well as the uncertainties of MCMC parameters,
\footnote{This is strictly true if the posterior distributions of all parameters (microlensing \& detrending parameters) are perfectly Gaussians. We have tested with examples that this assumption is valid under our parameterization.}
but can significantly reduce the dimensions of the parameter space for MCMC searches.

\subsection{Microlensing Source Color-color Relation $(K_p-I)$ vs. $(V-I)$} \label{sec:color-color}

We derive the relation between $K_p-I$ and $V-I$ that applies to the source stars of \K2C9 microlensing events. Here $K_p$ is the stellar magnitude in the Kepler bandpass. Such a color-color relation is important for multiple reasons. Technically, given the high noise level in \K2C9 data and the complex model we use, we would like to assess our results by comparing the derived $K_p-I$ colors with the expected values from this color-color relation. Scientifically, knowing the source flux in \emph{Kepler} bandpass can be crucial for constraining the parallax parameters for events that are extremely faint and/or do not peak within the \K2C9 window. This has been demonstrated in ground-based observations \citep{Yee:2012} and the \emph{Spitzer} microlensing campaigns \citep[e.g.,][]{SCN:2015,Zhu:2017}. It can also play a crucial role in breaking in the general four-fold degeneracy in the single-satellite experiment \citep{Refsdal:1966,Gould:1994}.

In previous studies, the color-color relations were always derived based on nearby ($<2'$) isolated stars. However, such an approach is not applicable to \K2C9 microlensing simply because there are not enough nearby stars that are isolated. 
Instead, we derive a theoretical color-color relation between $K_p-I$ and $V-I$, using the synthetic stellar spectra and known bandpass transmission functions. We choose $V$ and $I$ mostly because they are the primary band passes used in ground-based microlensing surveys. The advantage of this choice is that the combination of $V$ and $I$ can essentially cover the broad $K_p$ bandpass, making the derived color-color relation less sensitive to the details of the stellar spectrum and the interstellar extinction.

In principle, one can also derive this color-color relation by using stars that have been observed by \emph{Kepler}, for example, stars in the prime \emph{Kepler} field. Indeed, by starting from the \emph{Kepler} Input Catalog \citep{Brown:2011} and using the filter transformations between $(V,I)$ and $(g,r,i,z)$ given in \citet{Blanton:2007}, we find the following linear relation between $K_p-I$ and $V-I$ for $0.5<V-I<1.5$,
\begin{equation} \label{eqn:color-color-linear}
    K_p-I = 0.729(V-I) + 0.044\ .
\end{equation}
However, this relation may not apply to the microlensing sources in \K2C9 for several reasons. First, the \emph{Kepler} stars are preferentially solar-like (FGK main-sequence) stars, but the microlensing sources are generally evolved stars with possibly later stellar types. Second, the interstellar extinction is very different for these two stellar samples. Such a difference appears as both the total amount of extinction and the form of the extinction curve.

\begin{figure*}
\epsscale{1.}
\plotone{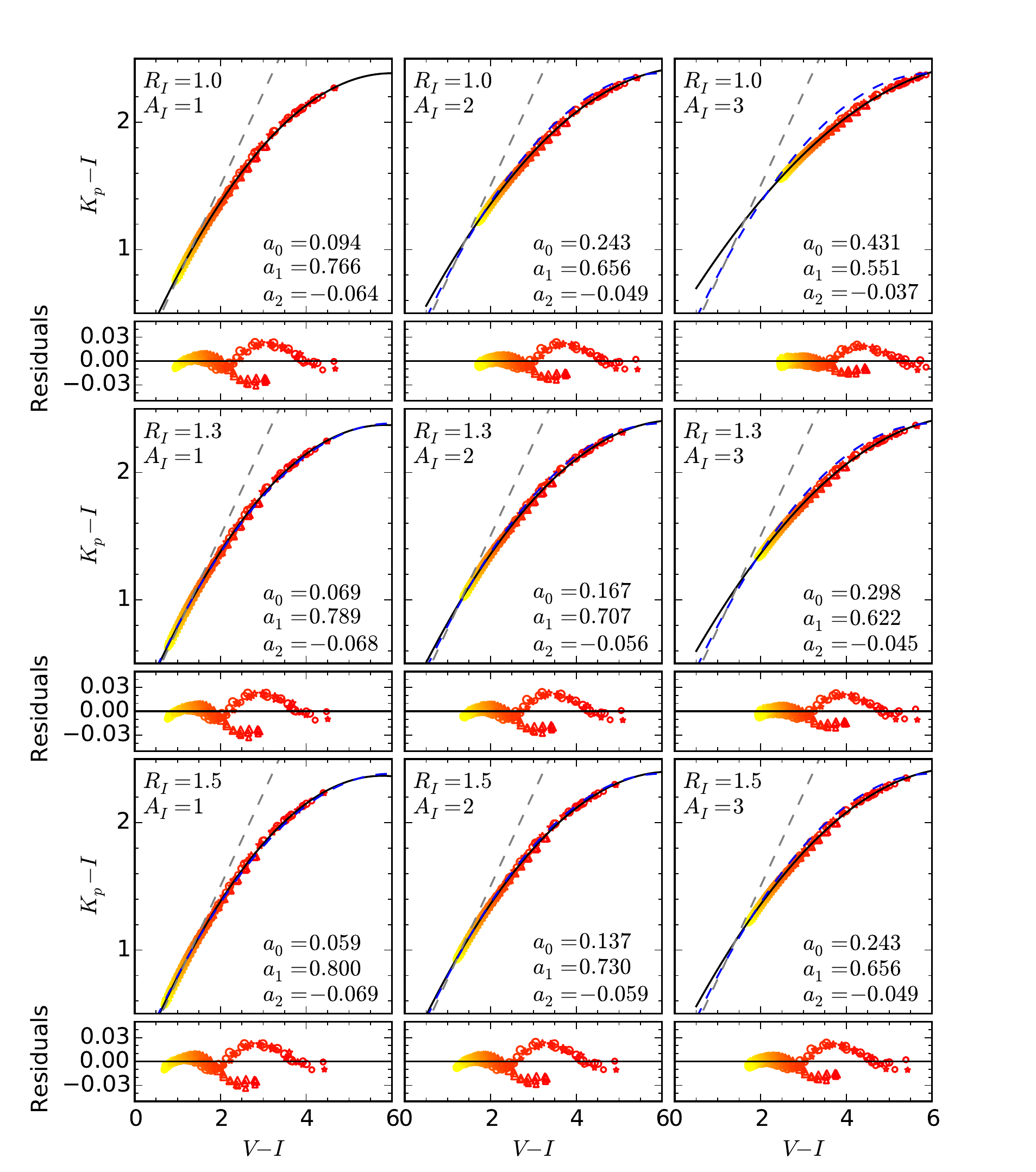}
\caption{The $K_p-I$ vs. $V-I$ color-color relations for different extinctions ($R_I$ \& $A_I$). Stellar colors derived from synthetic spectra are shown, with sizes standing for different surface gravity ($\log{g}=$2, 3, and 4), colors for the effective temperature (3000~K$\le T_{\rm eff}\le$8000~k), and shapes for metallicities (triangle for [Fe/H]=-2, asterisk for [Fe/H]=0, and circle for [Fe/H]=+0.5). The gray dashed line is the linear relation that is derived for FGK-type stars in the prime \emph{Kepler} field, which explains the color-color relation reasonably well for $V-I<1.5$. In each panel, the best quadratic fit to the color-color relation is shown as a solid black curve, with the coefficients given at the lower right corner. {As an illustration of the impact of the extinction, we also plot the quadratic fit from the top left panel as blue dashed curves in all other panels.}
\label{fig:colors}}
\end{figure*}

We therefore want to derive a color-color relation that better applies to microlensing stars. We start from the PHOENIX synthetic stellar spectra \citep{Husser:2013}, and use the $\{K_p,V,I\}$ wavelength response functions \citep{Bessell:2005,Handbook} to compute the stellar flux in each bandpass. This routine is applied to various stars with effective temperature between 3000~K and 8000~K, surface gravity $\log{g}=(2,3,4)$, and metallicity [Fe/H]$=(-2.0,0.0,+0.5)$. Then for chosen extinction parameters $A_I$ and $R_I$, the $K_p-I$ and $V-I$ colors can be derived. We show in Figure~\ref{fig:colors} all the synthetic stellar colors for typical extinction parameters, $R_I=(1.0,1.3,1.5)$ and $A_I=(1,2,3)$. Because for bulge stars, the extinction parameters can be known from red clump stars \citep{Nataf:2013}, the $K_p-I$ vs. $V-I$ color-color relations are provided for different combinations of $R_I$ and $A_I$\deleted{, and are illustrated in Figure~\ref{fig:colors}}.

As {Figure~\ref{fig:colors}} illustrates, the linear relation we derived from \emph{Kepler} stars (Equation~\ref{eqn:color-color-linear}) only apply to stars with $V-I\lesssim1.5$. For larger $V-I$ values, the $K_p-I$ color is better described as a quadratic function of $V-I$
\begin{equation} \label{eqn:color-color-quadratic}
    K_p-I = a_2(V-I)^2+a_1(V-I)+a_0\ .
\end{equation}
These coefficients are given in Figure~\ref{fig:colors} for different combinations of $R_I$ and $A_I$. As shown in the residual plots of Figure~\ref{fig:colors}, these quadratic color-color relations can predict the $K_p-I$ color to {within $\sim$0.02~mag for a broad range of $V-I$ colors. The remaining uncertainty is primarily due to metallicity effect.}
Such a precision is more than adequate for our purpose, {because the source $V-I$ color measured from the ground-based data has a typical uncertainty $\gtrsim$0.05~mag}.

%%%%%%%%%%%%%%%%%%%%%%%%%%%%%%%%%
\section{Applications to Known Microlensing Events} \label{sec:examples}

We use two microlensing events to demonstrate the ability of our method in measuring the microlensing parallax parameter. These events were found and observed by the fourth phase of the Optical Gravitational Lensing Experiment (OGLE-IV, \citealt{Udalski:2003,Udalski:2015b}) using its 1.3 m Warsaw Telescope at the Las Campanas Observatory in Chile. These events do not show any anomalous features due to lens companions, based on the reasonably good coverage from OGLE.
\footnote{It is still possible that there can be anomalous features in \K2C9 data, because the space-based and ground-based light curves are sensitive to different regions in the planetary parameter space. See \citet{Poleski:2016b} for an example. However, this is unlikely in most cases, based on various other constraints. See \citet{Zhu:2017} and \citet{Chung:2017} for cases of OGLE-2015-BLG-0961 and OGLE-2015-BLG-1482, respectively.}
Therefore, the microlensing model of these events only involves single-lens parameters $\{t_0,u_0,t_\e,\rho,\piee,\pien\}$.

For each event, the \K2C9 light curve that has the minimum scattering after detrending is used. We model the space-based and ground-based data simultaneously following the method in Section~\ref{sec:modeling}. The modeling is performed for different values of $N_\detr$, for purposes to check the stability of our method and the consistency with expectations. For the latter, we specifically compare the $K_p-I$ color that is given by the best-fit model and the one that is predicted by the color-color relation. The calibration of the OGLE-IV photometric data to the standard system is then required, and this is done by following the method of \citet{Udalski:2015b}, which is accurate to 0.02~mag \citep{Udalski:2008,Szymanski:2011}. For the \emph{Kepler} data, we use zero point ZP=25 to convert between flux values and $K_p$ magnitudes.

\subsection{OGLE-2016-BLG-0980: A Faint Event with Quiet Background}

\begin{figure*}[ht!]
\epsscale{1.2}
\plotone{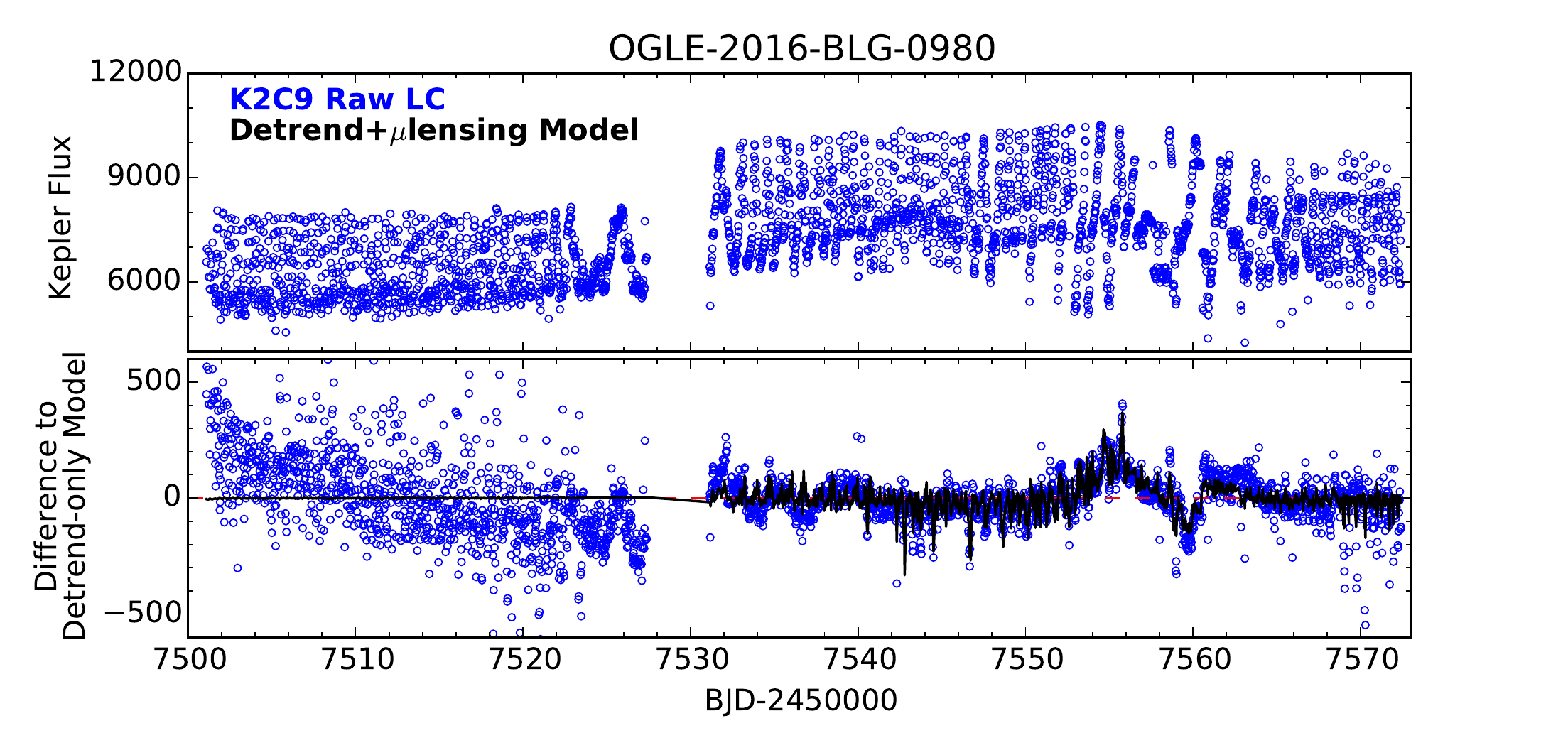}
\caption{The \K2C9 raw light curve (blue) and best-fit models of OGLE-2016-BLG-0980. Here the ``Detrend-only model'' (red) is the best-fit detrend model without the microlensing signal (i.e., Equation~\ref{eqn:model-kepler} without the first term), and the ``Detrend+$\mu$lensing model'' (black) is the one that simultaneously fits the microlensing signal and the detrending parameters (i.e., Equation~\ref{eqn:model-kepler}). The microlensing signal is not visible in the raw light curve, but the ``Detrend+$\mu$lensing'' model leads to a detection of the signal with $\Delta \chi^2=560$ compared to the ``Detrend-only'' model. Note that the data from C9a are affected by unknown systematics, but this does not affect modeling the microlensing signal.
\label{fig:0980-kepler}}
\end{figure*}

\begin{figure}
\epsscale{1.2}
\plotone{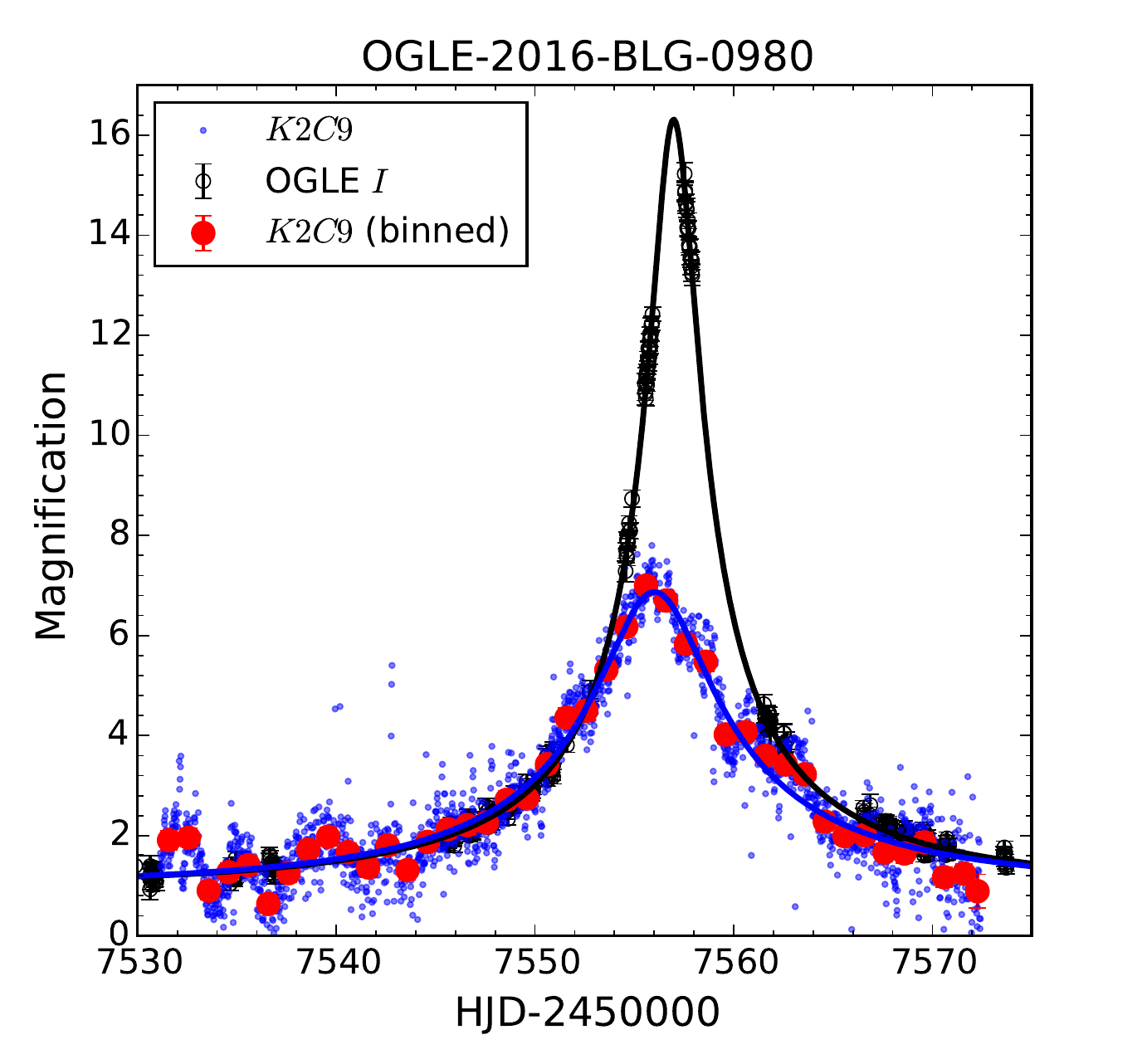}
\caption{Light curves of OGLE-2016-BLG-0980\ in views of OGLE (black) and \emph{Kepler}. For \emph{Kepler}, we show the original data (blue dots) and the 1-day binned data (red dots with error bars). The best-fit microlensing models are also shown in solid curves. The difference between the space-based and ground-based light curves {gives a measurement of} the microlensing parallax effect.
\label{fig:0980-joint}}
\end{figure}

\begin{figure}
\epsscale{1.2}
\plotone{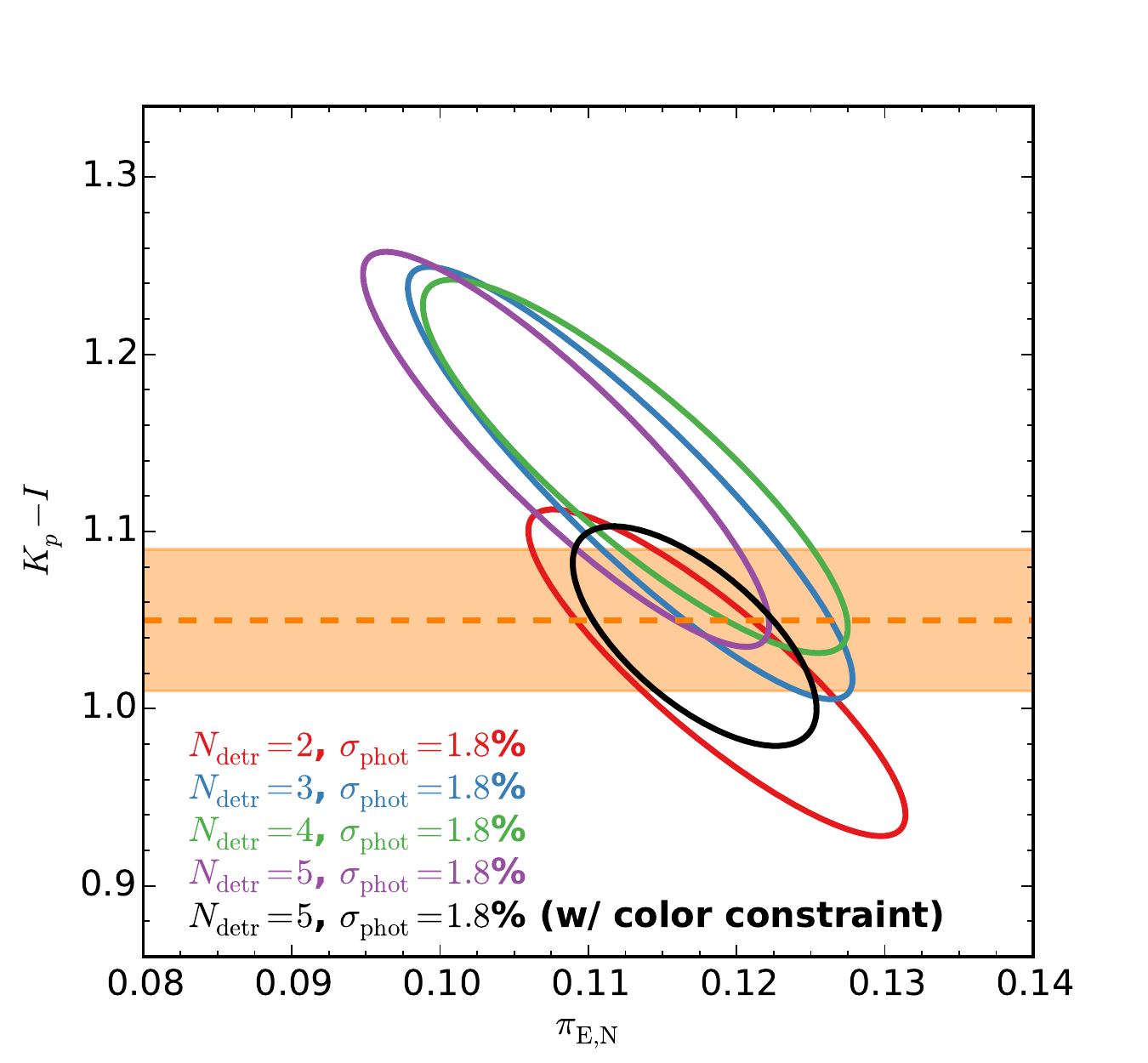}
\caption{The $1-\sigma$ constraints on the source $K_p-I$ color and the $\pien$ parameter for event OGLE-2016-BLG-0980. Different panels correspond to results from the photometry with different aperture size $R_{\rm apt}$. In each panel, the ellipses are the derived $1-\sigma$ error contours for modelings with different number of detrending terms $N_\detr$, and the orange band represents the range of $K_p-I$ that is predicted by the source $V-I$ color and the color-color relation in Section~\ref{sec:color-color}.
\label{fig:0980-check}}
\end{figure}

\begin{deluxetable}{llll}
\tablecaption{Best-fit parameters and associated uncertainties of different fits for event OGLE-2016-BLG-0980. For the last one, we require $(K_p-I)_\s=1.05\pm0.04$ and allow for a 3-$\sigma$ range.
\label{tab:0980}}
\tablehead{\colhead{Parameters} & \colhead{OGLE-only} & \colhead{OGLE+\emph{K2}} & \colhead{OGLE+\emph{K2}} \\
\colhead{} & \colhead{} & \colhead{} & \colhead{(color constraint)}}
\startdata
$t_0-2450000$ & 7556.976(6) & 7556.979(5) & 7556.979(5) \\
$u_0$ & 0.0591(22) & 0.0616(15) & 0.0608(15) \\
$t_\e$ (days) & 21.0(5) & 20.5(5) & 20.6(4) \\
$\pien$ & \nodata & 0.114(14) & 0.116(8) \\
$\piee$ & \nodata & 0.056(9) & 0.055(6) \\
$(K_p-I)_\s$ & \nodata & 1.15(11) & 1.04(5) \\
\enddata
\end{deluxetable}

Event OGLE-2016-BLG-0980
\footnote{This is the same event as OGLE-2016-BLG-1020. See the OGLE event pages at \url{http://ogle.astrouw.edu.pl/ogle4/ews/2016/blg-0980.html} and \url{http://ogle.astrouw.edu.pl/ogle4/ews/2016/blg-1020.html}.}
has a typical timescale ($t_\e=21~$days) and a relatively faint source ($I_\s=19.004\pm0.036$) that has a typical color [$(V-I)_\s=1.425\pm0.036$]. 
%Here values of $I_\s$ and $(V-I)_\s$ are after the calibration to OGLE-III using Equation~(\ref{eqn:calibration}). 
According to \citet{Nataf:2013}, the extinction parameters toward this event are found to be $A_I=0.90$ and $R_I=1.23$. {With these extinction parameters and by interpolating} the $(K_p-I)$ vs. $(V-I)$ color-color relations given in Figure~\ref{fig:colors}, we expect the source $K_p-I$ color to be $1.05\pm0.04$.

We model the \K2C9 data and OGLE data simultaneously following the method given in Section~\ref{sec:methodology}, and provide the parameters with associated uncertainties in Table~\ref{tab:0980}. As a demonstration of the method rather than a detailed characterization of the event, we only focus on one out of the four degenerate solutions. Figure~\ref{fig:0980-kepler} shows the raw \K2C9 light curve of OGLE-2016-BLG-0980. The microlensing signal is not noticeable in this raw light curve. However, after applying the best-fit detrend-only model (i.e., a model given by Equation~\ref{eqn:model-kepler} but without the first term), we find a significant trend centered at JD$'\equiv$JD$-2450000=7556$ in the residuals, which is close to the expected peak of the microlensing event. Then we simultaneously model the detrending terms and the microlensing signal (i.e., the model given by Equation~\ref{eqn:model-kepler}). We refer this model as the ``detrend+$\mu$lensing model'', and assign the uncertainty to individual data point such that the $\chi^2$ per degree of freedom is unity. We re-scale the $\chi^2$ from the detrend-only model with this normalized uncertainty, and find the detrend+$\mu$lensing model can fit the data better than the detrend-only model by $\Delta\chi^2=560$.
\footnote{We refer to the fit with $N_\detr=5$.}
As a demonstration of the final light curve product, we show in Figure~\ref{fig:0980-joint} the space-based and ground-based light curves, in which we only preserve the microlensing signal in the \K2C9 raw data.

The best-fit parameters as well as the associated uncertainties are derived for different values of $N_\detr$. The results are shown in Figure~\ref{fig:0980-check} in the $(K_p-I)$ vs. $\pien$ plane, along with the $(K_p-I)$ value that is predicted from the color-color relation. In each case, we also report the photometric precision, which is defined as the ratio between the level of scattering (root-mean-square, or rms) in residuals and the median flux.

As shown in Figure~\ref{fig:0980-check}, the resulting constraints on $\pien$ and $K_p-I$ are fairly stable regardless of the choice of $N_\detr$. The derived $(K_p-I)$ colors also agree with the expected value reasonably well. The total flux (target+background) is equivalent to a $K_p=15$ point source, and our method can lead to a photometric precision of $1.8\%$. For a comparison, the microlensing source has $K_p=20$.

It is also suggested by this event that, for the source $(K_p-I)$ color, the derived uncertainty (under no constraint) is usually larger than the uncertainty inferred from the color-color relation. Therefore, by imposing a prior on the source $(K_p-I)$ color, we should be able to improve the final constraint on $\pien$, given the strong correlation between these two. For example, we use $N_\detr=5$ and allow $(K_p-I)$ to vary within the 3-$\sigma$ range of the expected value, we are able to reduce the uncertainty on $\pien$ by $40\%$. Although this does not seem particularly important for the current event, the color constraint can play a crucial role in constraining the parallax parameters for events with extremely faint baseline or events with partial light curve coverages \citep[e.g.,][]{SCN:2015,Zhu:2017,Shvartzvald:2017}.

\subsection{OGLE-2016-BLG-0940: A Bright Event with Noisy Background}

\begin{figure*}
\epsscale{1.2}
\plotone{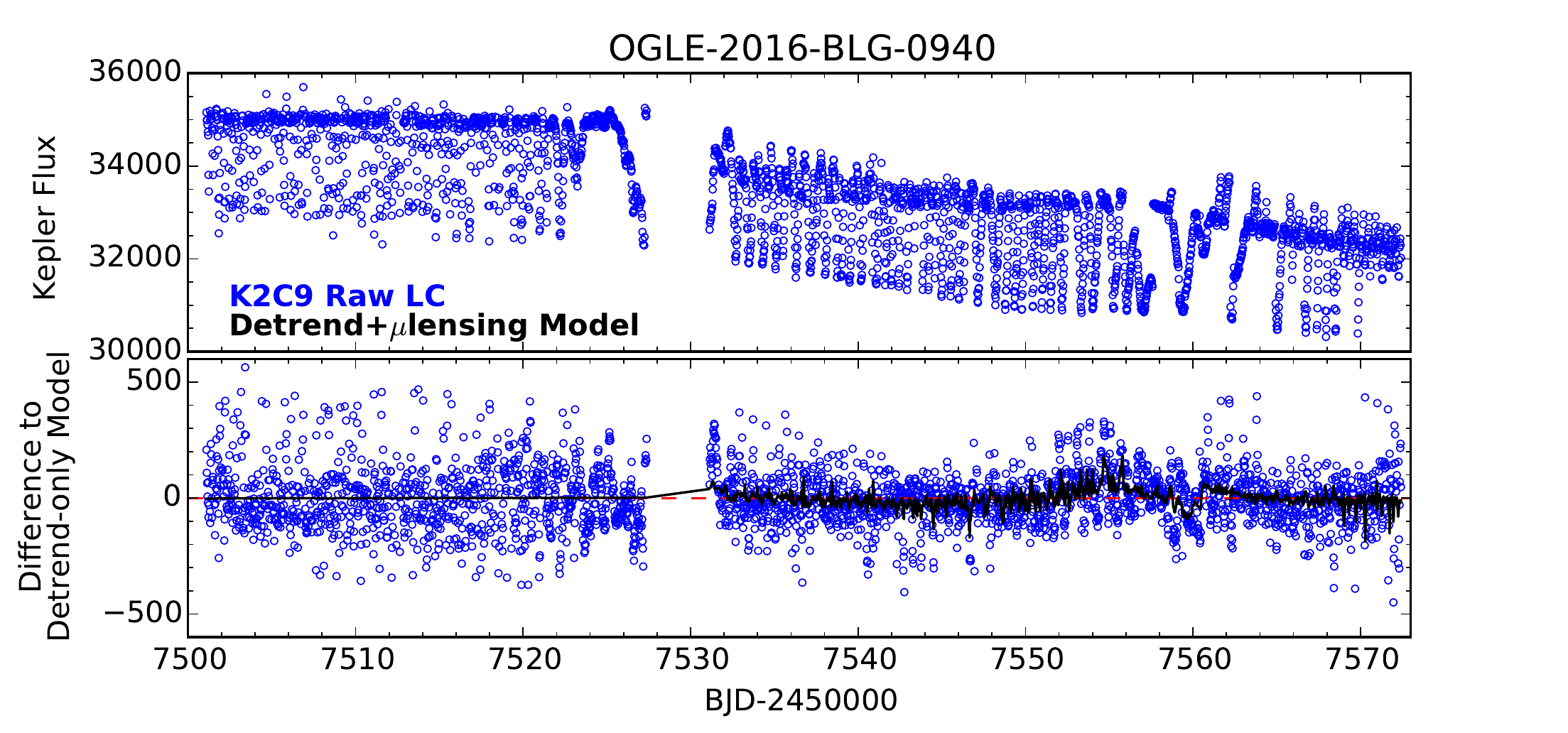}
\caption{Similar to Figure~\ref{fig:0980-kepler}, but for event OGLE-2016-BLG-0940. The microlensing signal is detected with $\Delta\chi^2=250$.
\label{fig:0940-kepler}}
\end{figure*}

\begin{figure}
\epsscale{1.2}
\plotone{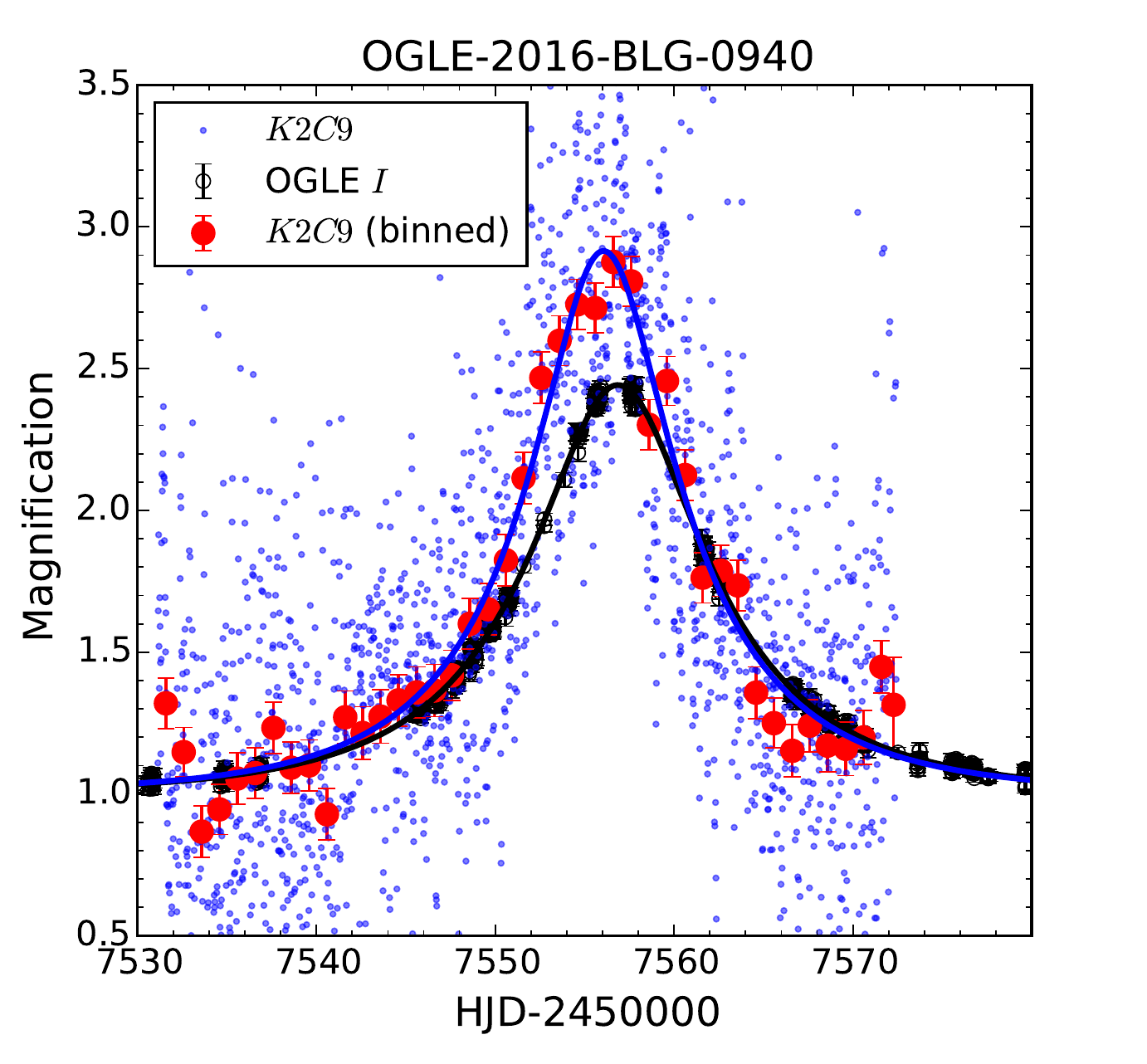}
\caption{Similar to Figure~\ref{fig:0980-joint}, but for event OGLE-2016-BLG-0940.
\label{fig:0940-joint}}
\end{figure}

\begin{figure}
\epsscale{1.2}
\plotone{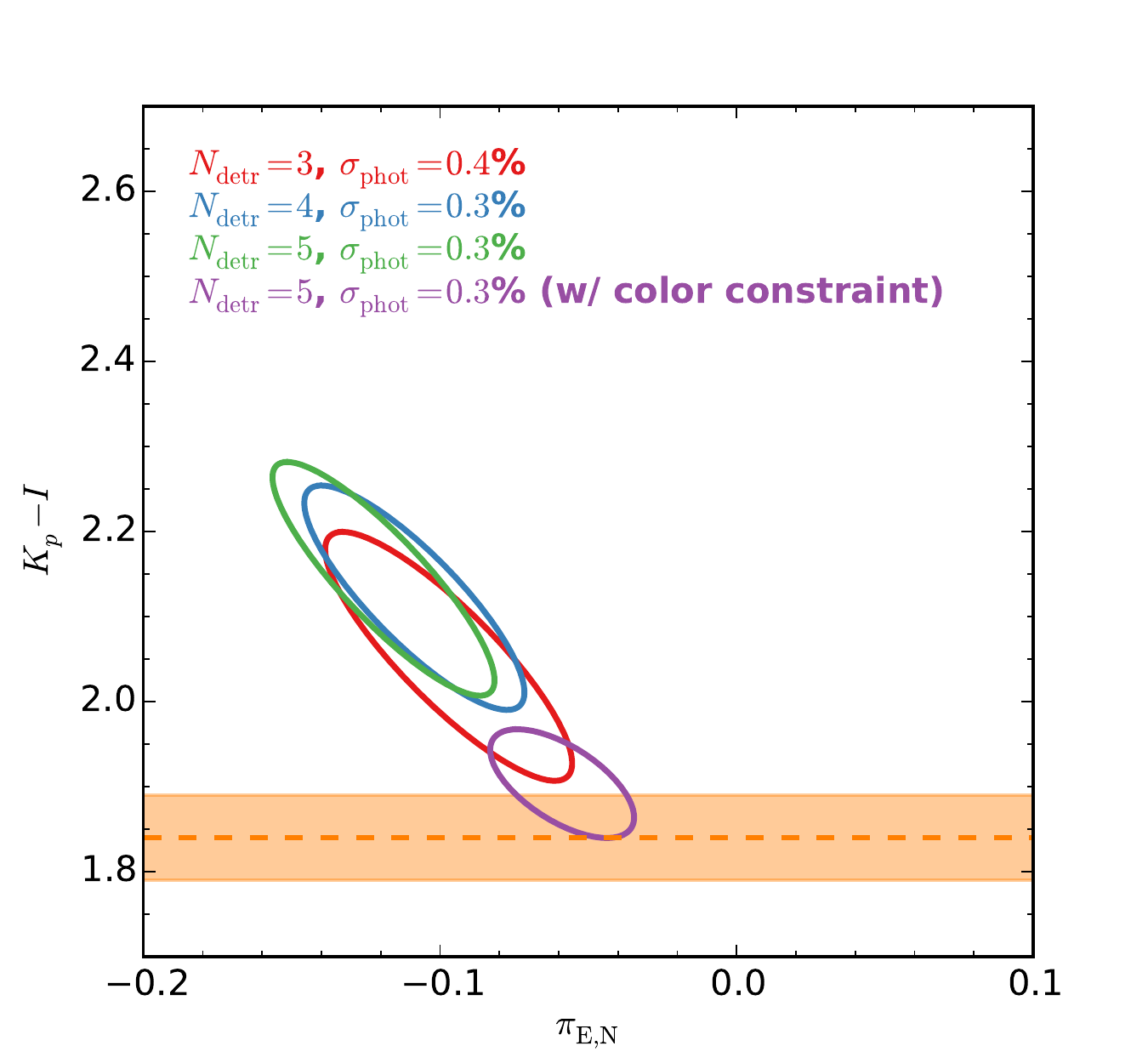}
\caption{The $1-\sigma$ constraints on the source $K_p-I$ color and the $\pien$ parameter for event OGLE-2016-BLG-0940. See the caption of Figure~\ref{fig:0980-check} for more details. In the current case, we only show the results with $N_\detr\ge3$.
\label{fig:0940-check}}
\end{figure}

\begin{deluxetable}{llll}
\tablecaption{Best-fit parameters and associated uncertainties of different fits for event OGLE-2016-BLG-0940. For the last one, we require $(K_p-I)_\s=1.84\pm0.05$ and allow for a 3-$\sigma$ range.
\label{tab:0940}}
\tablehead{\colhead{Parameters} & \colhead{OGLE-only} & \colhead{OGLE+\emph{K2}} & \colhead{OGLE+\emph{K2}} \\
\colhead{} & \colhead{} & \colhead{} & \colhead{(color constraint)}}
\startdata
$t_0-2450000$ & 7556.855(13) & 7556.858(14) & 7556.854(13) \\
$u_0$ & 0.434(11) & 0.431(6) & 0.428(6) \\
$t_\e$ (days) & 11.43(18) & 11.48(11) & 11.52(10) \\
$\pien$ & \nodata & $-0.119(37)$ & -0.059(24) \\
$\piee$ & \nodata & 0.106(24) & 0.098(18) \\
$(K_p-I)_\s$ & \nodata & 2.14(14) & 1.90(6) \\
\enddata
\end{deluxetable}

Event OGLE-2016-BLG-0940
\footnote{\url{http://ogle.astrouw.edu.pl/ogle4/ews/2016/blg-0940.html}.}
is relatively bright ($I_\s=17.3$), and has a timescale $t_\e=11$ days. Compared to the previous one, this event has brighter and noisier background in \K2C9 due to several bright neighboring stars. This event also suffers from more extinction, with $R_I=1.23$ and $A_I=2.6$ \citep{Nataf:2013}, and thus the source star appears redder. With $(V-I)_\s=3.14\pm0.06$, we expect that the source has $(K_p-I)_\s=1.84\pm0.05$.

The raw \K2C9 light curve of OGLE-2016-BLG-0940 is shown in Figure~\ref{fig:0940-kepler}. Similar to the previous example, the (distorted) microlensing signal can only be marginally seen after the detrend-only model is applied. A simultaneous modeling of the systematics and the microlensing signal leads to the improvement of the fit by $\Delta\chi^2=250$. The cleaned \K2C9 light curve is shown in Figure~\ref{fig:0940-joint}, together with the OGLE-IV light curve.

We also derive the best-fit parameters and their associated uncertainties for different choices of $N_\detr$, and show the results in terms of constraints in the $(K_p-I)$ vs. $\pien$ plane in Figure~\ref{fig:0940-check}. Compared to event OGLE-2016-BLG-0980, the uncertainties we derive for OGLE-2016-BLG-0940 on $K_p-I$ are larger, even though the current event has a brighter source. This is because of the stronger systematic effect in the present event.  Nevertheless, models with $N_\detr\ge3$ give results that are consistent with each other, and the derived $K_p-I$ colors are also consistent with the value predicted by the color-color relation within 2-$\sigma$ level, as seen in Figure~\ref{fig:0940-check}. Once the color constraint is imposed, the uncertainties on $\pien$ and $(K_p-I)$ color are reduced significantly, {and the new measurements are consistent within 2-$\sigma$ of the ones without color constraint}. See Table~\ref{tab:0940} for the best-fit parameters and associated uncertainties of the modelings with only OGLE data, OGLE+\K2C9, and OGLE+\K2C9 together with $(K_p-I)$ constrained by the color-color relation.

Because of the brighter aperture, the photometric precision we can achieve for this event, 0.4\%, is better than that we obtain for OGLE-2016-BLG-0980, but the scattering in the residuals remains at the similar level (130 compared to 120 for OGLE-2016-BLG-0980, both in the \emph{Kepler} flux unit). 

%%%%%%%%%%%%%%%%%%%%%%%%%%%%%%%%%
\section{Discussion} \label{sec:discussion}

In this work, we present the method that can be used to extract microlensing signals from the \emph{K2} Campaign 9 data {for known events}. Our method relies on the photometric reduction technique that is introduced in \citet{MSF:2017}, and the key component in this technique is the derivation of global astrometric solutions of individual \emph{K2} frames \citep{Huang:2015}. We then introduce the method to measure the microlensing parameters in the raw \K2C9 light curve, which is to fit the systematic trend and the microlensing signal simultaneously. We also provide analyses of two known OGLE-IV microlensing events as applications of this method.

We derive the $K_p-I$ color vs. $V-I$ color relation based on synthetic stellar spectra and $(K_p,V,I)$ transmission curves. With this color-color relation, we are able to predict the microlensing source flux in the $K_p$ bandpass to $\lesssim3\%$, based on the known source $V-I$ color from ground-based observations. We use the predicted $K_p-I$ color to validate the result of the light curve modeling, and find reasonable agreements between these two for both example events. Furthermore, given that the $K_p-I$ vs. $V-I$ color-color relation is fairly precise and invariant to the properties of stars, and also that the \emph{K2} data are affected by strong systematics, we suggest that this relation should be applied to improve the light curve modeling whenever possible.

For an estimation of the signal-to-noise ratio (S/N) of the microlensing signal, we use
\begin{equation} \label{eqn:snr}
    {\rm S/N} = \alpha \frac{(A_\max-1)F_\s^\kep}{\sigma_\phot \langle F_{\rm tot}^\kep \rangle} \sqrt{\frac{t_\e}{T_{\rm int}}}\ ,
\end{equation}
where $A_\max$ is the maximum magnification as seen by \emph{Kepler}, $T_{\rm int}\equiv30~$min is the integration time of \emph{K2} observations, and $\langle F_{\rm tot}^\kep \rangle$ is the averaged total flux within the \emph{K2} aperture. For single-lens events, we find the coefficient $\alpha\approx3.5$ based on the two events in Section~\ref{sec:examples}. In addition, we find that for the (microlensing and non-microlensing) targets we examined, the product $\sigma_\phot \langle F_{\rm tot}^\kep \rangle$, which is the level of scattering in the cleaned data, is $\sim150$. This is equivalent to a single measurement with 100\% uncertainty on a $K_p=19.6$ star, or a typical deblended microlensing source [$I_\s=18$ and $(V-I)_\s=2.5$]. Equation~\ref{eqn:snr} is useful for assessing the detectability of a given event, as well as estimating the completeness of our method in detecting microlensing signals.

The photometric technique that is used in this work uses a large number ($\sim$1000) of bright point-like sources to derive the astrometric solutions, and thus is not directly applicable to targets outside the super stamp region. Therefore, additional techniques \citep[e.g.,][]{Pal:2016} are required in order to extract the raw light curves of $\sim$70 \K2C9 late targets \citep{Henderson:2016} and 33 microlensing events in the \emph{K2} Campaign 11. Nevertheless, our method for interpreting the \emph{K2} light curves is applicable to microlensing events both inside and outside the super stamp region.

Although the events we analyzed in this paper do not have the finite-source effect, it is common for events that show planetary or binary anomalies \citep{Zhu:2014,Suzuki:2016}. The characteristic timescale of the finite-source effect is the source radius crossing time
\begin{equation}
    t_\star = \frac{\theta_\star}{\mu_\rel} \approx 60~{\rm min} \left(\frac{\theta_\star}{0.6~\microas}\right) \left(\frac{\mu_\rel}{5\masyr}\right)^{-1}\ .
\end{equation}
It is comparable to the integration time of \emph{K2} long cadence observations, especially for events with lenses in the galactic disk. This has to be taken into account when the finite-source effect is present.

\acknowledgements
We thank Andy Gould and Scott Gaudi for discussions. {We also thank the anonymous referee for useful comments which helped to improve the manuscript.} Work by W.Z. was supported by US NSF grant AST-1516842. 
R.P. acknowledges support from K2 Guest Observer program under NASA grant NNX17AF72G.
This paper includes data collected by the Kepler mission. Funding for the Kepler mission is provided by the NASA Science Mission directorate.
Some of the data presented in this paper were obtained from the Mikulski Archive for Space Telescopes (MAST). STScI is operated by the Association of Universities for Research in Astronomy, Inc., under NASA contract NAS5-26555. Support for MAST for non-HST data is provided by the NASA Office of Space Science via grant NNX09AF08G and by other grants and contracts.
OGLE project has received funding from the National Science Centre, Poland, grant MAESTRO 2014/14/A/ST9/00121 to AU.

\end{CJK*}
\end{document}